\input{aipcheck}

\documentclass[
    ,final            
  ]
  {aipproc}

\layoutstyle{8x11single}

\def\sss{\scriptscriptstyle}
\def\^#1{^{\sss #1}}
\def\_#1{_{\sss #1}}
\def\beq{\begin{equation}}
\def\eeqno#1{\label{#1}\end{equation}}

\def\rarrow{\rightarrow }

\def\dleft{\rlap{{\it D}}\raise 8pt\hbox{$\scriptscriptstyle\Leftarrow$}}
\def\dright{\rlap{{\it D}}\raise 8pt\hbox{$\scriptscriptstyle\Rightarrow$}}

\def\cmss{{\rm cm~s^{-2}}}

\def\msun{M\_{\odot}}
\def\az{a\_{0}}

\def\baz{\bar a_0}

\def\l0{\ell\_{0}}

\def\rar{\rightarrow}

\def\l{\lambda}

\def\f{\phi}

\def\k{\kappa}

\def\r{\rho}

\def\m{\mu}

\def\n{\nu}

\def\F{\mathcal{F}}
\def\L{\mathcal{L}}

\def\Q{\mathcal{Q}}

\def\D{\Delta}

\def\d{\delta}

\def\drt{d^3\vr}
\def\a{\alpha}
\def\b{\beta}
\def\c{\gamma}
\def\d{\delta}
\def\xlimin{{x\rarrow\infty \atop{\raise 1pt
\hbox to 30pt{\rightarrowfill}}}}
\def\limlim#1#2{{#1\rarrow #2 \atop{\raise 1pt
\hbox to 30pt{\rightarrowfill}}}}

\def\vr{{\bf r}}

\def\vv{{\bf v}}

\def\vg{{\bf g}}

\def\grad{\vec\nabla}
\def\div{\vec \nabla\cdot}
\def\gf{\grad\phi}

\def\vgt{\tilde\vg}
\def\ft{\tilde\f}
\def\gft{\grad\tilde\f}
\def\fpg{4\pi G}

\def\aL{a\_{\Lambda}}
\def\aH{a\_H}
\def\sss{\scriptscriptstyle}
\def\^#1{^{\sss #1}}
\def\_#1{_{\sss #1}}
\def\gmn{g\_{\m\n}}
\def\Gmn{g\^{\m\n}}
\def\hgmn{\hat g\_{\m\n}}
\def\ten#1#2{^{\sss#1}_{\sss#2}}
\def\C{\Gamma}
\def\hC{\hat\C}
\def\Up{\Upsilon}
\def\hGmn{\hat \Gmn}

\def\hgmn{\hat g\_{\m\n}}
\def\hgh{\hat g^{1/2}}
\def\gh{g^{1/2}}

\def\gft{\grad\tilde\f}
\def\fh{\hat\f}
\def\gfh{\grad\fh}
\def\M{\mathcal{M}}

\begin{document}
\title{New Physics at Low Accelerations (MOND):
an Alternative to Dark Matter} 

\author{Mordehai Milgrom}{
  address={The Weizmann Institute Center for Astrophysics,
   Rehovot 76100, Israel}}

\begin{abstract}
I describe the MOND paradigm, which posits a departure from standard
physics below a certain acceleration scale. This acceleration as
deduced from the dynamics in galaxies is found mysteriously to agree
with the cosmic acceleration scales defined by the present day
expansion rate and by the density of `dark energy'. I put special
emphasis on phenomenology and on critical comparison with the
competing paradigm based on classical dynamics plus cold dark
matter. I also describe briefly nonrelativistic and relativistic
MOND theories.

\end{abstract}

\maketitle

{\it ``There can be no doubt that the interplanetary and
interstellar spaces are not empty but are occupied by a material
substance, or body, which is certainly the largest, and probably
most uniform, body of which we have any knowledge.'' }(J. C.
Maxwell).

{\it ``En astronomie, nous voyons les corps dont nous \'etudions les
mouvements, et nous admettons le plus souvent qu'ils ne subissent
pas l'action d'autres corps invisibles.''}  (H. Poincar\'e).
\section{introduction}
Normal matter in the universe makes itself felt in many ways: It
interacts through strong and electromagnet interactions; so it emits
and absorbs radiation. It self collides efficiently under
astrophysical conditions, and dissipates energy. It produces
magnetic fields that can be felt by particle acceleration, or by
inducing synchrotron emission, for example. It decays and produces
fast particles. It causes mechanical effects as in supernova
explosions, or in ram-pressure stripping of gas in galaxies. In
contradistinction, the putative, omnipresent dark matter (DM) is not
required (and is not known) to do any of these things. The only
trait it is required to share with standard matter is gravity: it is
invoked only so as to modify and enhance the gravitational field of
the visible baryons. It is thus conceivable that DM does not, in
fact, exist, and that the extra gravity it is purported to supply is
provided by a modification of the standard laws of dynamics.
\par
So, following Poincar\'e's admonition, and not Maxwell's misguided
conviction, we have been pursuing for a quarter of a century now a
new paradigm of dynamics, dubbed `MOND'. Its main raison d'etre
remains the explanation of dynamics in the universe without the need
for DM, which by and large, it does very well (with some
exceptions). In particular, right at its advent it made a number of
strong predictions, which have been confirmed over the years, and
unearthed a number of unsuspected regularities in the properties of
galactic systems. But, an additional motivation for embracing MOND
has emerged with the recognition that the constant characterizing
the paradigm may have cosmological origin, and that MOND as
expressed in local physics has symmetries that may have descended
from cosmology.

\section{The MOND paradigm}
MOND starts by introducing into physics a new constant, $\az$, with
the dimensions of acceleration. This constant marks the borderline
between the pre-MOND physic--valid approximately for accelerations
much larger than $\az$--and the MOND regime of much lower
accelerations, just as $\hbar$ does in the context of quantum
physics, and $c$ in that of relativity. The MOND constant also
enters strongly into physics in the MOND regime and is predicted to
appear in many apparently unconnected laws and relations in this
regime of phenomena. To recover standard physics in the high
acceleration regime, we require that when we formally take $\az\rar
0$ in all the equations of motion, a MOND theory tends to standard
physics (like the restoration of classical predictions when
$\hbar\rar 0$ in quantum predictions for various observables).
Finally, to achieve the phenomenological goals of MOND, we require
that in the deep-MOND limit of {\it nonrelativistic} (NR) physics;
i.e., formally taking $\az\rar \infty$, and the masses $m\rar 0$
so that $m\az$ is fixed,
the theory becomes scale
invariant \cite{mil09a}; i.e., invariant under
$(t,\vr)\rar\l(t,\vr)$. To see what this means, take, for example, a
system of masses $m_i$ with all accelerations much smaller than
$\az$. Then, if $\vr_i(t)$ describes some possible history of the
system, $\l\vr_i(t/\l)$ is also a solution of the theory. For
example, the length of the planetary year in a deep-MOND planetary
system is proportional to the orbital size, not to its $3/2$ power
as in the Newtonian Kepler's 3rd law.
\par
Look, indeed, at the quintessential problem of the circular motion
of a test particle at a distance $R$ around a point mass $M$. On
dimensional grounds alone, the expression for MOND acceleration of
the particle must be of the form \beq a= {MG\over
R^2}\n\left({MG\over R^2\az}\right).\eeqno{muta} The MOND basic
tenets require that for $MGR^{-2}\gg\az$ we have $\n\rar 1$, while
for $MGR^{-2}\ll\az$ we have to have scale invariance, which
dictates $\n(y)\approx y^{-1/2}$ (the normalization is absorbed in
the definition of $\az$), or
  \beq a\approx (MG\az)^{1/2}/R.  \eeqno{upo}
\par
There are three main points to take home from this simple,
introductory case, which are also relevant to more general systems:
1. For a given mass, $M$, the asymptotic acceleration at large radii
goes as $R^{-1}$ compared with the standard $R^{-2}$ (accelerations
scale as $\l^{-1}$). 2. In this region, $a$ is proportional to
$M^{1/2}$, not to $M$. 3. The transition occurs always at the same
value of the acceleration (not a fixed radius).
\par
 Extensive reviews of
various aspects of the MOND paradigm can be found in
\cite{sm02,scarpa06,bek06,mil08,skordis09}.
\section{MOND theories}
\par
The above basic tenets of MOND can be incorporated into various MOND
theories. For example, in the NR regime one can modify Newtonian
gravity by generalizing the Poisson equation for the gravitational
field \cite{bm84,mil09d}. Another option, which cannot be described
as modified gravity is to modify the kinetic action of particles
leading to modification of Newton's second law, or to modified
inertia \cite{mil94}.

\subsection{Nonrelativistic theories}
The action governing a Newtonian system made of a gravitating mass
distribution $\r(\vr)$, which produces a gravitational potential
$\f$, can be written as $I=\int (\L\_K+\L\_P) \drt dt$, with
$\L\_K=\r\vv^2/2$ ($\vv$ is the velocity field) and $\L\_P=-(8\pi
G)^{-1}(\gf)^2-(1/2)\r\f$. Bekenstein and Milgrom \cite{bm84}
generalized this Lagrangian by replacing $(\gf)^2$ with
$\az^2\F[(\gf)^2/\az^2]$ to obtain a MOND generalization of the
Poisson equation
  \beq \div[\m(|\gf|/\az)\gf]=4\pi G\r,  \eeqno{poisson}
with $\m(x)=\F'(x^2)$ satisfying $\m(x\gg 1)\approx 1,~\m(x\ll
1)\approx x$. This theory had been, for many years, the only
complete MOND theory in stock. It has been applied extensively to
many problems, both analytically and numerically. For example, solar
system tests \cite{mil86a,bm06,mil09c}, forces on massive bodies
\cite{mil97,mil02,dms08}, disc stability and bar formation
\cite{brada99,tiret08}, two-body relaxation \cite{cb04}, dynamical
friction \cite{nip08}, escape speed from a galaxy \cite{fbzh07,
wu07}, galaxy interactions (Combes and Tiret, this volume,
\cite{tiret07, nlc07a}), galaxy collapse \cite{nlc07b}, triaxial
models of galactic systems \cite{wang08, wu09}, the external-field
effect as applied to dwarf spheroidals and warp induction
\cite{brada00a,brada00b,angus08}, structure formation (e.g.,
\cite{ll08}), and quite a few more.
\par
This formulation of MOND has also been interpreted as resulting for
the omnipresence of a gravitationally polarizable medium
\cite{bl07}, with a relativistic generalization \cite{blt08,blt09}.
\par
It has been realized recently \cite{mil09d} that a larger family of
modified-gravity MOND theories exist, which involve two or even more
potentials, with only one of them coupling to matter. For example,
start from a Palatini-type formulation of Newtonian gravity,
introducing beside $\f$ an acceleration field $\vgt$, and taking
instead of the above Poissonian Lagrangian density:
 \beq \L\_P={1\over 8\pi
G}(\vgt^2+2\gf\cdot\vgt)-{1\over 2}\r\f.  \eeqno{iiisut}
 It gives,
upon variation over $\vgt$ and $\f$: $\vgt=-\gf$, and
$\div\vgt=-\fpg\r$, respectively, which is indeed standard Newtonian
gravity. If we generalize this action by replacing $\vgt^2$ with
$\az^2\Q[\vgt^2/\az^2]$ we get a theory that is equivalent to that
described by eq.(\ref{poisson}). However, if we require a priori
that $\vgt$ is derivable from an auxiliary potential $\vgt=-\gft$,
we get a new theory that is a quasi-linear MOND (QUMOND) theory,
which is rather easier to apply because it requires solving only
linear differential equations \cite{mil09d}
 \beq \Delta\ft=\fpg\r,
 ~~~~ \Delta\f=\div[\n(|\gft|/\az)\gft],
 \eeqno{poisa}
with $\n(y)\equiv\Q'(y^2)$. This theory requires solving the Poisson
equation twice, with a nonlinear algebraic step in between. It has
been used recently, beside eq.(\ref{poisson}), to calculate MOND
effects in the inner solar system \cite{mil09c}.
\par
We can get even a larger class of theories by making $\L\_P$ a
function of the three scalars $(\gf)^2,~(\gft)^2, ~\gf\cdot\gft$
(and possibly involving even more potentials) \cite{mil09d}. Look,
for example, at actions of the form
  \beq \L=-{1\over 8\pi
G}\{\b(\gf)^2+\a(\gfh)^2-\az^2\M[(\gf-\gfh)^2/\az^2]\}
  +\r({1\over 2}\vv^2-\f),  \eeqno{futcol}
which leads to the field equations
  \beq \div[\m^*(|\gft|/\az)\gft]=\fpg\r,
  ~~~~~~\D\f=\div[(1-\a^{-1}\M')\gft], \eeqno{hutred}
  with
   \beq \ft\equiv \f-\fh,~~~~~\m^*=
   \b-{\a+\b\over\a}\M'[(\gft/\az)^2]. \eeqno{cutes}
The first equation in (\ref{hutred}) is solved for $\ft$, and then
the second is a Poisson equation for the MOND potential $\f$, with
the known right hand side as source. The parameter range $0<\b+\a\le
1$ is excluded. The limiting case $\b=-\a=1$  is particularly
simple, as the theory then reduces to the QUMOND theory
(\ref{poisa}) with $\n(y)=1+\M'(y^2)$. These theories have lead to a
class of relativistic, bimetric MOND (BIMOND) theories (see below).

\subsubsection{Modified inertia}
An altogether different route to constructing MOND theories(e.g.,
\cite{mil94,mil08}) is to replace the kinetic action $\sum_i
m_i\vv^2_i/2$ (written now for a collection of point masses) by a
more general functional of the trajectories. Such theories have,
generically, to be time nonlocal if they are to obey Galilei
invariance and the proper Newtonian and MOND limits \cite{mil94}.
\par
Interestingly, such theories predict a universal equation that
determines the orbital motion on circular trajectories in
axisymmetric potentials, relating the orbital speed, $V$, and
radius, $R$, by
 \beq\m(V^2/R\az)V^2/R=-{\partial\f\_N\over\partial R},  \eeqno{gulba}
where $\f\_N$ is the {\it Newtonian} potential, and $\mu(x)$ is a
function that derives from the action of the theory as specialized
to circular, constant-speed orbits. This relation is unique for a
given theory (i.e., system independent). All MOND rotation-curve
analyses to date employ this relation [not the relation derived from
theories such as eq.(\ref{poisson}) or eq.(\ref{poisa}), which would
require a new numerical calculation for each system analyzed].
\par
Such `modified-inertia' theories can differ greatly from
modified-gravity formulations of MOND concerning some phenomena,
even in the NR regime. Whereas in the latter, the anomalous MOND
acceleration of test particles in the field of a given mass depends
only on the position in the (modified) field, the time nonlocality
of the former theory can produce retardation and hysteresis effects,
and makes the anomalous acceleration at any location depend on
properties of the whole orbit. For example, if the accelerations are
small on some segments of a trajectory, MOND effects can be felt
also on segments where the accelerations are high. This can give
rise, for example, to different MOND effects on bound and unbound
orbits, or on circular and highly elliptic orbits. In the solar
system this can differentiate between planets on one hand, and long
period comets and unbound spacecraft, such as the Pioneers, on the
other (Milgrom, in preparation).
\par
Also, in such theories the predicted behavior of different MOND
effects, such as the external-field effect (EFE), can be rather
different from that in modified-gravity theories (Milgrom, in
preparation).

\subsection{Relativistic theories}
One naturally wants to incorporate the MOND principles in a
relativistic extension. The most studied theory following this
effort is Bekenstein's Tensor-Vector-Scalar (TeVeS) theory
\cite{bek04}, which is an outgrowth of Sanders's stratified gravity
approach \cite{sanders97}. There is a large body of work elaborating
on, extending, reinterpreting, and criticizing TeVeS. See, e.g.,
\cite{bek06,zfs06,zfs07,bruneton07,contaldi08,sagi08,sagi09}, and,
in particular, see the extensive reviews \cite{bek06,skordis09}.
\par
There is also initial investigation of relativistic bimetric MOND
(BIMOND) theories, brought to light recently\cite{mil09e}, which I
describe succinctly below.
\par
All such theories, even if they turn out to work well
phenomenologically, must be only effective, approximate theories as
evinced by the appearance of a free function in them, which is to
reproduce the interpolating function in equations such as
eq.(\ref{muta}). The exact form of this function will hopefully some
day emerge from a deeper theory underlying the effective ones.

\subsubsection{TeVeS}
Here are the main features of this much discussed theory.
\begin{itemize}
\item Gravity in TeVeS \cite{bek04}
is described by a metric $ g_{\alpha\beta}$, as in General
Relativity (GR), plus a vector field, ${\cal U}_\alpha$, and a
scalar field $\phi$. (In other formulations the scalar is eliminated
\cite{zfs06}.
\item Matter is coupled to one combination of
the fields: the physical metric $\tilde g_{\alpha\beta}\equiv
e^{-2\phi}(g_{\alpha\beta} + {\cal U}_\alpha {\cal U}_\beta) -
e^{2\phi} {\cal U}_\alpha {\cal U}_\beta$.
\item $g_{\alpha\beta}$ is governed by the usual Hilbert-Einstein
action,
the vector field (constrained to have a unit length) by a
Maxwell-like action,\footnote{This was later modified to
obviate some inconsistencies in the original version \cite{skordis08}.}
and the scalar action can be written as $ S_s
=-{ 1\over 2Gk\hat k}\int Q\big[\hat k (g^{\alpha\beta} -{\cal
U}^\alpha {\cal U}^\beta)\phi_{,\alpha}\phi_{,\beta} \big](-g)^{1/2}
d^4 x. $
\item There are three constants: $k, ~\hat k$, and a parameter
$K$ appearing in the vector action, and one free
function $Q(x)$, which engenders the interpolating function of MOND
in the NR limit.
\item For NR systems TeVeS reproduces the NR
MOND phenomenology with the MOND potential being the sum of two
potentials, one satisfying the (linear) Poisson equation, the other
satisfying eq.(\ref{poisson}) with $\az\propto k\hat k^{-1/2}$.
\item For weak fields ($\f\ll c^2$) TeVeS gives lensing
according to the standard GR formula  but with the MOND potential.
\item Structure formation and CMB: According to preliminary work,
TeVeS has the potential to mimic aspects of cosmological DM
\cite{sanders05,dodelson06,skordis06,skordisetal06,zfs08}.
\end{itemize}

\subsubsection{Bimetric MOND gravity}
Inspired by bi-potential theories governed by the Lagrangian density
(\ref{futcol}), the BIMOND theories \cite{mil09e} involve two
metrics as independent degrees of freedom. One, $\gmn$, is the MOND
metric, which alone appears in the matter action, and which couples
in the standard way to matter. The other metric, $\hgmn$, is an
auxiliary one. In constructing an action for the theory we now have
at our disposal the usual scalars made of the curvature tensors of
the two metrics such as the Ricci scalars $R$ and $\hat R$. But, in
addition we can construct scalars using the tensor difference
between the Levi-Civita connections of the two metrics:
 \beq  C\ten{\a}{\b\c}=\C\ten{\a}{\b\c}-\hC\ten{\a}{\b\c}.
  \eeqno{veyo}
This is particularly germane in the context of MOND: Connections,
and hence the $C\ten{\a}{\b\c}$, play the role of gravitational
accelerations. So, without introducing new constants in the
relativistic context we can use the MOND constant $\az$ to form
dimensionless tensors, $\az^{-1}C\ten{\a}{\b\c}$, and from these
construct scalars.\footnote{I work in units where $c=1$, otherwise
we use the MOND scale-length $\ell=c^2/\az$ in the dimensionless
tensors $\ell C\ten{\a}{\b\c}$.} Functions of these scalar can then
be used as Lagrangian densities.
\par
Examples of scalars that are quadratic in $C\ten{\a}{\b\c}$ are
 \beq\Up^{(2)}_i=\Gmn C\ten{\c}{\m\l}C\ten{\l}{\n\c},
 ~~~\bar C\^{\c}C\_{\c},~~~
 \gmn\bar C\^{\m}\bar C\^{\n},~~~\Gmn C\_{\m} C\_{\n},
 ~~~g\_{\a\l}g\^{\b\m} g\^{\c\n}C\ten{\a}{\b\c}C\ten{\l}{\m\n},
  \eeqno{kurta}
where $\bar C\^{\c}\equiv \Gmn C\ten{\c}{\m\n},~C\_{\c}\equiv
C\ten{\n}{\c\n}$, as well as others, e.g., contractions with
$\hgmn$, in addition to higher, $m$ powers $\Up^{(m)}_i$. We thus
seek to construct a covariant Lagrangian density that is a function
of $R,~\hat R$, and scalars $\Up_i$. I take the Lagrangian to be
linear in the curvature scalars, so as not to end up with a higher
derivative theory. The tensor $C\ten{\a}{\b\c}$ and the scalars
$\Up_i$ contain only first derivatives of the metrics, so we can
take functions of them that interpolate between the MOND
($\Up^{(2)}_i/\az^2\ll 1$) and standard ($\Up^{(2)}_i/\az^2\gg 1$)
regimes.
\par
I have thus considered \cite{mil09e} actions of the form
   \beq I=-{1\over 16\pi G}\int[\b\gh R + \a\hgh \hat R
 -(g\hat g)^{1/4}f(\k)\az^2\M(\Up/\az^2)]d^4x
 +I\_M(\gmn,\psi_i)+\hat I\_M(\hgmn,\chi_i),  \eeqno{gedap}
where for brevity's sake I write $\M$ as a function of only one
quadratic scalar. Its derivative $\M'$ plays the role of an
interpolating function between the MOND and conventional regime.
Also, $I\_M$ is the matter action, with matter degrees of freedom
represented by $\psi_i$, coupling only to $\gmn$ [$g$ and $\hat g$
are minus the determinants of the two metrics, and $\k=(g/\hat
g)^{1/4}$]. I also permit twin matter (TM) described by degrees of
freedom $\chi_i$, which couples only to $\hgmn$.
\par
We see that the modification of GR entailed by MOND does not enter
here by modifying the `elasticity' of space-time (except perhaps its
strength), as is done in $f(R)$ theories and the like. The
modification is introduced through the interaction between the
space-time on which `our' $\psi$ matter lives and the auxiliary one.
In a membrane description of gravity, we can say that matter lives
on one membrane of a pair, with the two membranes, each with its own
standard elasticity, coupled together. The way the shape of the home
membrane is affected by matter then depends on the combined
elasticity properties of the double membrane. However, matter
response depends only on the shape of its home membrane. Such
heuristics may, in fact, lead to a fundamental understanding of the
origins of the MOND paradigm, and the meaning of the length
$\ell=c^2/\az$, which enters NR physics as $\az$.
\par
In particular, I found scalar arguments of $\M$ constructed from the
tensor $\Up\_{\m\n}=C\ten{\c}{\m\l}C\ten{\l}{\n\c}
-C\ten{\c}{\m\n}C\ten{\l}{\l\c}$ to lead to simple theories. For
example $\Gmn\Up\_{\m\n}$, or $\hGmn\Up\_{\m\n}$ (or a symmetric
combination of both). Such a tensor is constructed from
$C\ten{\a}{\b\c}$ in the same way as the first-derivative part in
the Ricci tensor is constructed from the connections. Also, the
subclass of theories with $\b+\a=0$, and arbitrary scalar argument,
is interesting and simple, and has been studied in more detail.
\par
Such theories have the following properties:
\begin{itemize}
\item Beside Newton's $G$ (and $c$) they involves only $\az$ as a
new constant.
\item
The equations of motion involve no higher than second derivatives in
the metrics.
\item
The nonrelativistic limit: On a locally double-Minkowski background,
the first order NR metric for any $\a,~\b$ is given by
$\gmn=\eta\_{\m\n}-2\f\d\_{\m\n}$, where $\f$ is the MOND potential
that is determined by the theory given by eq.(\ref{hutred}), and
$\hgmn=\eta\_{\m\n}-2\fh\d\_{\m\n}$, with $\f^*=\f-\fh$ being the
Newtonian potential. QUMOND is gotten for $\b=-\a=1$. I also
considered backgrounds other than double Minkowski.
\item
Gravitational lensing by slowly moving masses: This has always been
a holy grail for relativistic MOND theories. It is thus reassuring
that the BIMOND theories predict enhanced, MOND-like lensing. Since
the relation between the full first order MOND metric and the MOND
potential is the same as the relation between the GR metric and the
Newtonian potential, the MOND potential controls lensing exactly as
the Newtonian potential does in GR. In other words, lensing and
massive-particle analysis (assuming GR) of the gravitational field
of a NR mass (e.g., a galaxy) would give the same MOND potential.
With choices of the scalar argument $\Up$ different from the above
this result can change somewhat; but, lensing is still enhanced and
MOND-like, i.e., underpinned by an asymptotic logarithmic potential
$\propto (MG\az)^{1/2}ln(r)$.
\item
The GR limit: We can choose the form of $\M$ so that in the formal
limit $\az\rar 0$ we obtain GR exactly, possibly with a `dark
energy' term of order $\M(\infty)\az^2$. For instance, in the case
$\b=1$ (and $G$ the Newton constant) the requirement is that
$\M'(z)\rar 0$ for $z\rar\infty$. This limit can be approached as
fast as required; so departures from GR in the inner solar system
and short period binary pulsars can be made as small as desired.
\item
Pinpointing the correct cosmology is still moot as it requires
additional assumptions on initial conditions, symmetries, matter
content (especially that of the twin matter), etc.. There exist,
however, interesting cosmological solutions. For example, with a
certain symmetry between the two types of matter there are
cosmologies with $\hgmn=\gmn$ with $\gmn$ describing a standard GR
cosmology (for $G/\b$ the Newton constant) with a cosmological constant
$\Lambda=-\M(0)\az^2/2(\a+\b)$. Such cosmologies  may be described
with the two `membranes' being stuck together ($\hgmn=\gmn$) on
large scales, while they separate locally ($\hgmn\not=\gmn$) due to
matter (and twin matter) inhomegenieties.
\par
More generally, there appears in these theories a weakly variable
`dark energy' term that is of order $\az^2\M$, which is of order
$\az^2$ if $\M$ is of order unity. For example, a de Sitter, or Anti
de Sitter, universe is a generic vacuum solution of these theories
with cosmological constant $\sim\M(0)\az^2$.
\par
By and large, cosmological considerations seem to prefer a
matter-twin-matter symmetric universe, with $\a=\b=1$. In this case
$C\ten{\a}{\b\c}=0$ in cosmology, and the cosmological equations
reduce to those of GR with a cosmological constant.
\par
Thus even without any connection with MOND, such theories may
provide alternatives to `dark energy', to be investigated alongside
existing schemes, such as $f(R)$ theories.
\item
The theory has the (yet unproven) potential to account for all
aspects of the `dark sector' (galactic `DM', cosmological `DM', and
`dark energy') in one fell swoop, with all controlled by $\az$.
\item
All vacuum solutions $\gmn$ of GR [with CC $\sim\az\^2\M(0)$] are
also vacuum solutions of BIMOND with $\hgmn=\gmn$. In particular, GR
gravitational waves, in vacuum, are also BIMOND solutions. Also, a
double Schwarzschild geometry is a vacuum solution of BIMOND
(corresponding to an equal matter-TM central masses).
\end{itemize}
\par
Another avenue for exploration brought to mind by these theories
starts from the interesting possibility that twin matter exists, and
then interacts `gravitationally' with normal matter, indirectly
through the $\M$ term.\footnote{It then does not play the role of
the putative dark matter; this role is still played by the MOND
departure from GR.} Preliminary study \cite{mil10a} shows this
effective interaction to be different from standard gravity, and
rather peculiar, in ways that depend on the BIMOND version at hand.
For example, in versions that are completely symmetric in the two
types of matter, gravity within each sector is described by standard
MOND, but matter and twin matter repel each other. The repulsion is
MOND like in the MOND regime but the interaction disappears
altogether at high accelerations. This would have obvious
ramifications for structure formation, for example, as it would
lead, among other things, to matter-twin-matter segregation.
\par
Be all this promising as it may, it represents only initial work,
and there are still important aspects of the theory to be checked.
These include questions of causality, existence of ghosts, etc..
Bimetric theories have a long history and there is a considerable
body of related literature discussing, among other things, the above
questions of principle for various classes of bimetric theories.
(See, for example, the recent treatment in
\cite{bdg06,bdg07,banados09}, where the authors consider bimetric
coupling involving the metrics themselves, not their derivatives, as
in BIMOND theories.) In particular, it was shown \cite{boulanger01}
that bimetric theories with derivative coupling, satisfying some
general assumptions, suffer generically from the existence of
ghosts. It remains to be seen, however, how relevant and/or
deleterious such results are to BIMOND, especially since a major
assumption underlying them does not apply in BIMOND.

\section{MOND phenomenology}
A MOND theory should predict the motions of test and non-test bodies
in the field of an arbitrary mass distribution, such as a galaxy.
For example, rotation curve analysis of disc galaxies partly tests
such predictions, as it probes the accelerations only in the
symmetry plane of the disc  and only for nearly circular orbits.
However, beyond such tests of MOND predictions for specific,
individual systems, MOND predicts a considerable number of relations
that should hold between galaxy properties, and which are the
analogous, in the realm of the galaxies, to Kepler's laws of
planetary motions. In such MOND laws, $\az$ appears in several
roles, similar to the appearance of $\hbar$ in many quantum
phenomena, such as the black-body spectrum, the photoelectric
effect, atomic spectra, superconductivity, etc., or the appearance
of the speed of light in relativistic phenomena, such as black-hole
physics, time dilation, or the relation between velocity and
momentum. Without the unifying force of the underlying theory the
appearance of the same constant in disparate, and apparently
unrelated, phenomena does not make sense.

\subsection{MOND laws of galactic motion}
\begin{figure}
\begin{tabular}{rl}
\tabularnewline
\includegraphics[width=0.8\columnwidth]{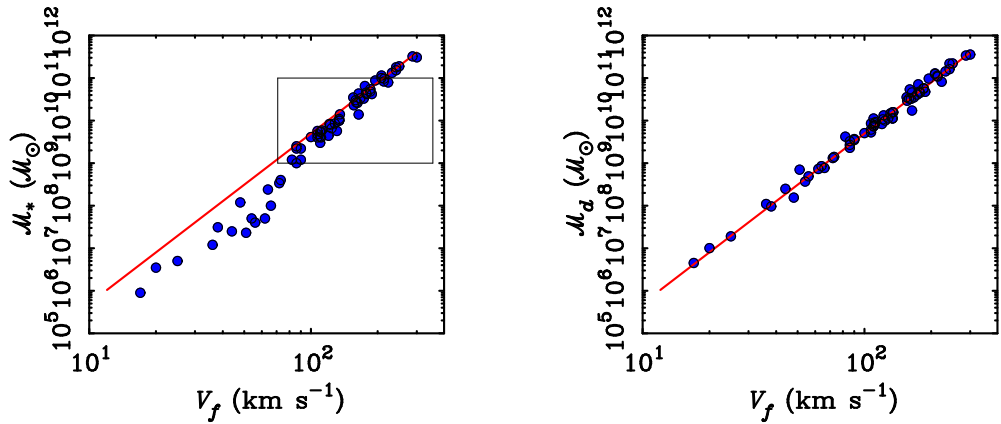} \\
\end{tabular}\par
\caption{Galaxy mass plotted against the rotation curve plateau
velocity. Left: analog of the traditional Tully-Fisher plot with
mass in stars only. Right: The total mass including that of gas. The
solid line has the log-log slope of 4, predicted by MOND, and is not
a fit \cite{mcg05} (the small rectangle shows where past analysis
had concentrated).} \label{fig1}
\end{figure}

I now list briefly some of these predicted MOND laws of galactic
motions. These, along with several others, are discussed, explained,
and referenced in more detail in other publications on MOND (e.g.
\cite{mil08}).
\begin{itemize}
\item
The rotational speed around an isolated mass becomes constant at
large radii (asymptotic flatness of rotation curves): $V(r)\rar
V_{\infty}$.
\item $V_{\infty}$ depends only on the total (`baryonic')
mass of the body via $V_{\infty}^4=MG\az$. This predicted the so
called `baryonic Tully-Fisher relation'
\cite{mcgaugh00,stark09,trachternach09} (see Fig.\ref{fig1} ).
\item  For quasi-isothermal systems, such as elliptical galaxies, the
velocity dispersion depends only on the total mass (in
contradistinction with the Newtonian virial relation) via
$\sigma^4\sim
MG\az$. This underlies the Faber-Jackson relation for elliptical
galaxies.
\item The mass discrepancy (transition from baryon dominance
to `DM' dominance) appears always around $V^2/R=\az$.
\item Isothermal spheres (e.g., as models of ellipticals) have mean
surface densities $\bar\Sigma\leq \az/G$, as is well supported by
the data (see, e.g., Figs. 1, 2 in \cite{graham06}).
\item
The central surface density of phantom `dark halos' is $\approx \az/
2\pi G$ \cite{mil09b,gentile09}, in accordance with the findings in
\cite{donato09}.
\item Discs with $\bar\Sigma\leq \az/G$ have added stability.
\item  Disc galaxies should exhibit a distinct disc component
of `DM', with predicted properties, in addition to the extended,
spheroidal `DM' component \cite{mil01}, as deduced by
\cite{kalberla07}.
\item Interpreting MOND with `DM' will result in negative
DM densities
in some, well specified, locations.
\end{itemize}
Several important facts about these predictions have to be noted:
(i) They follow as inevitable consequences of the basic MOND tenets,
and are oblivious to the exact way(s) in which galaxies or other
galactic systems formed. (ii) They are independent predictions in
the sense that without MOND none of them follows from the
others.\footnote{It is possible to build families of galaxies of
baryons plus DM that will satisfy any subset of these laws, but not
the rest. So in the framework of DM they'll each require a separate
explanation.} (iii) Inasmuch as they have been tested they are well
consistent with the data. (iv) In the framework of DM such laws must
follow from very strict connection between the amount and
distribution of DM and those of the baryons, because they relate
baryonic properties--such as (baryonic) masses--to properties
determined mainly by the DM--such as speeds. (v) Even without
further theoretical development, or interpretation, MOND has already
directed the eye to many regularities not suspected before,
including the appearance of an acceleration constant, $\az$, in many
a priori unrelated facets of galaxy dynamics.
\par
I now discuss briefly some of these laws: The independence of speed
on orbital radius in the limit of large radii follows directly from
the scale invariance of the deep-MOND limit: under space-time
scaling sizes change, but velocities do not.
\par
The MOND mass-asymptotic-speed relation, underlies the Tully-Fisher
relation (TFR), whose existence was known before the advent of MOND.
However, the traditional TFR correlates some luminosity measure of
galaxies with some velocity measure, with different choices giving
different results. To my knowledge the TFR has no creditable
theoretical basis in the DM doctrine. MOND has specified exactly
what is to be correlated with what: the total baryonic mass with the
asymptotic, constant rotational speed. This has lead to the
so-called baryonic TF correlation, which indeed is very tight, and
which conforms exactly with the predicted MOND relation: the power
of 4 and the proportionality factor being $G\az$, well consistent
with other determinations of $\az$ \cite{mcgaugh00,mcg05,stark09}.
Figure \ref{fig1} shows the results of one such analysis.

\par
 The MOND constant, $\az$, defines a scale of mass surface density
$\az/G$. This is why several of the MOND laws predict a special
value of the surface density in different contexts. For example, one
of the above laws, which follows from detailed analysis, states that
isothermal spheres--which in MOND have a finite mass, unlike their
Newtonian analogs--cannot have mean surface densities much exceeding
$\az/G$. This simply reflects the fact the IS disobeying this
inequality have mean accelerations larger than $\az$, and so are
Newtonian, and so cannot exist as finite mass objects.
\par
This last MOND prediction assigns a special surface density to the
normal matter component of a galactic object. But there is also an
interesting MOND prediction that has come to light recently, which
pertains to a property of the pure, fictitious `DM halo' needed to
explain MOND results with DM. This prediction identifies $\az/2\pi
G$ as a special central surface density of such phantom halos that
is of universal significance \cite{mil09b,gentile09}. The discovery
of this prediction was incited by a recent finding to this effect
\cite{donato09}. The two predictions follow from completely
different aspects of MOND, but both come up with a special surface
density determined by $\az/G$, which is the only scale available in
MOND. The fact that the same value of $\az$ determined from other
predictions gives a very good agreement with the findings of
\cite{donato09}, can count as another success of MOND.

\subsection{Rotation Curves of Disc Galaxies}
The quintessential MOND achievement, however, is the prediction of
detailed rotation curves (RCs) of many individual galaxies, from
their baryon distribution alone. Interestingly, the first systematic
analysis of such predictions appeared only some four years after the
advent of MOND, for it had had to await the appearance of extended
HI rotation curves. The first systematic study \cite{kent87} was
followed by amending analysis in \cite{mil88}; then, many such
analyses followed; e.g.,
\cite{bbs91,sanders96,sv98,mcgaugh98,dbm98,bottema02,
begum04,gentile04,gentile07a,corbelli07,barnes07,sn07,ms07}.

\begin{figure}
\begin{tabular}{lll}
\includegraphics[width=0.36\columnwidth]{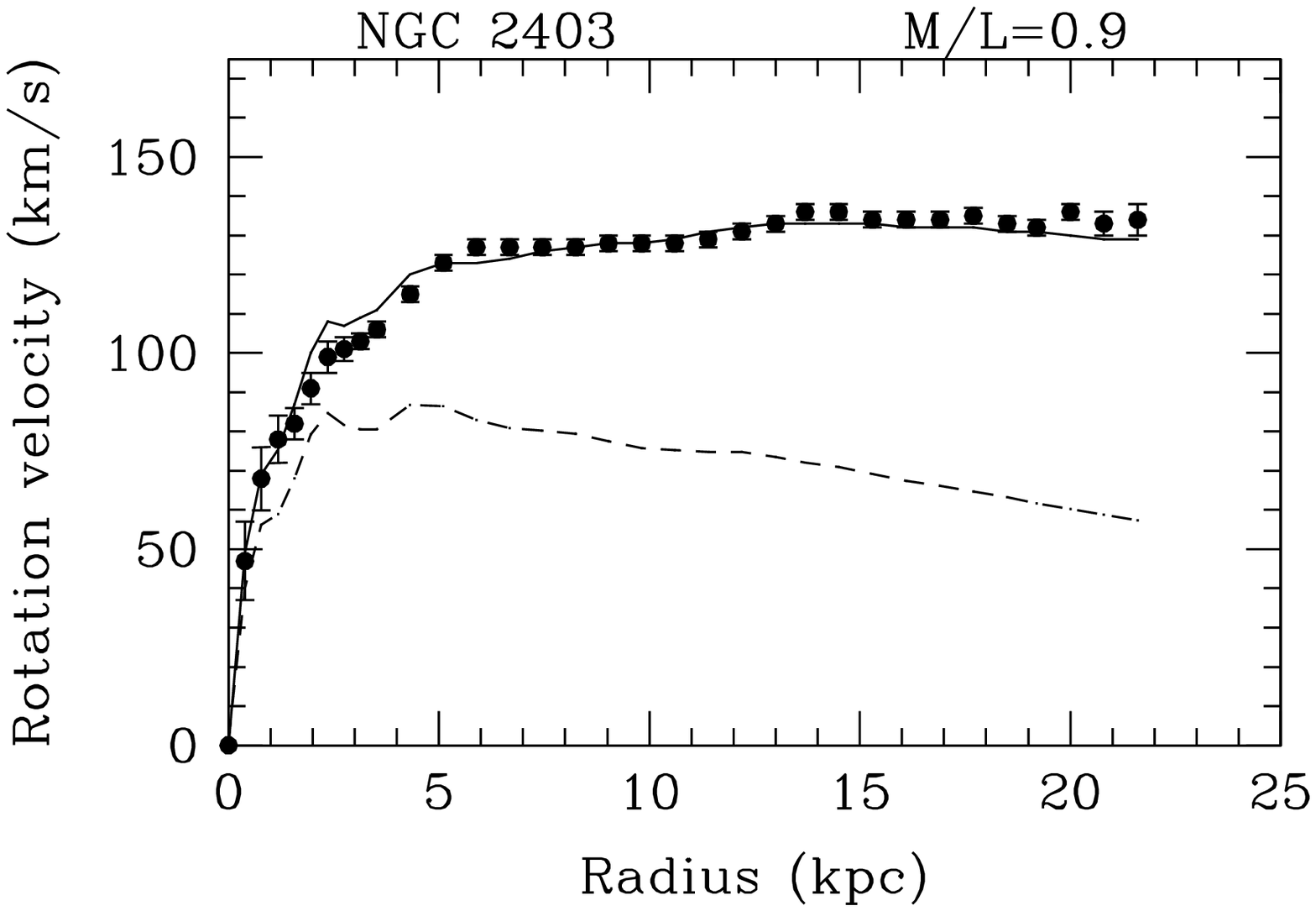}&
\includegraphics[width=0.32\columnwidth]{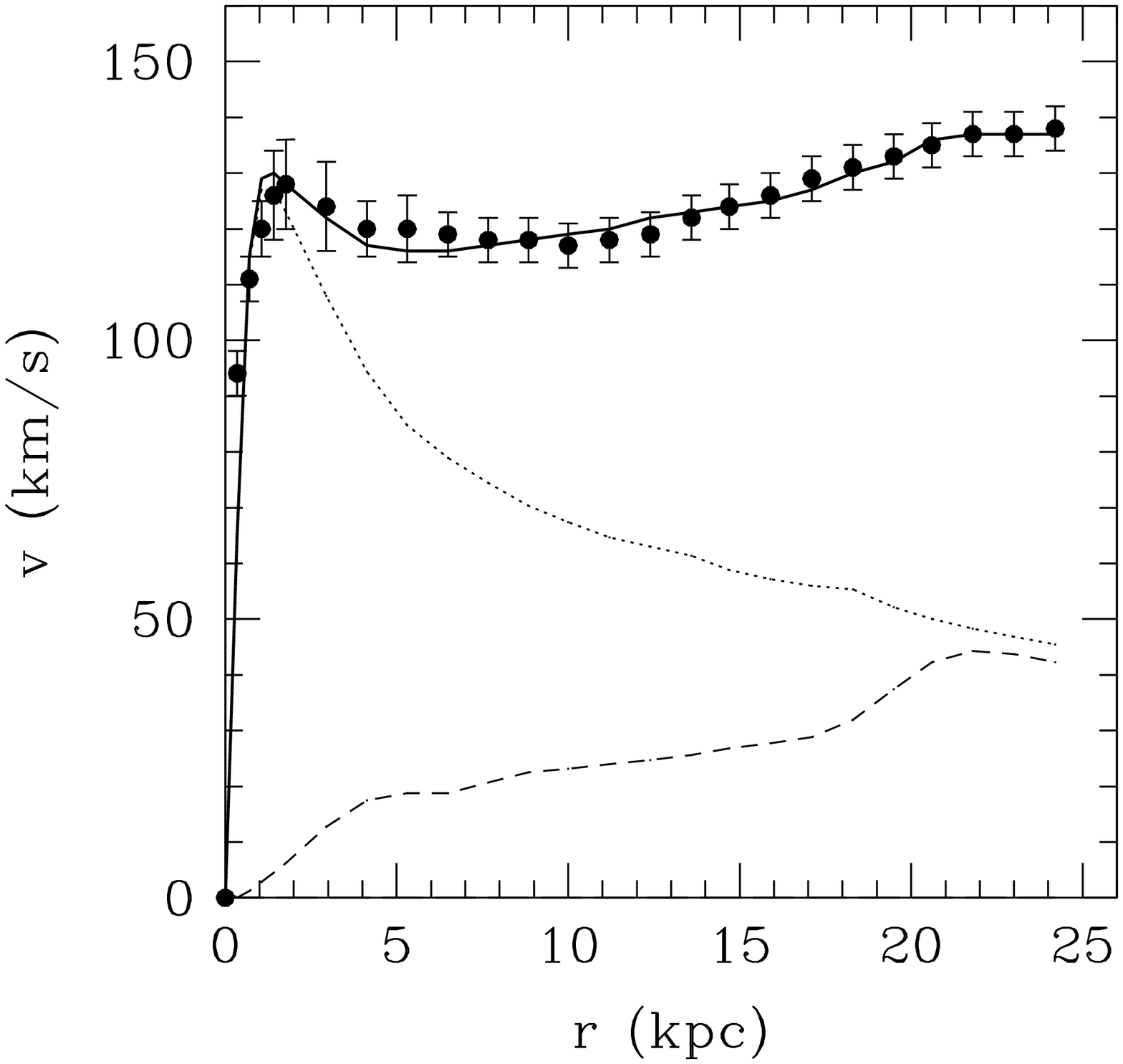}\\
\includegraphics[width=0.65\columnwidth]{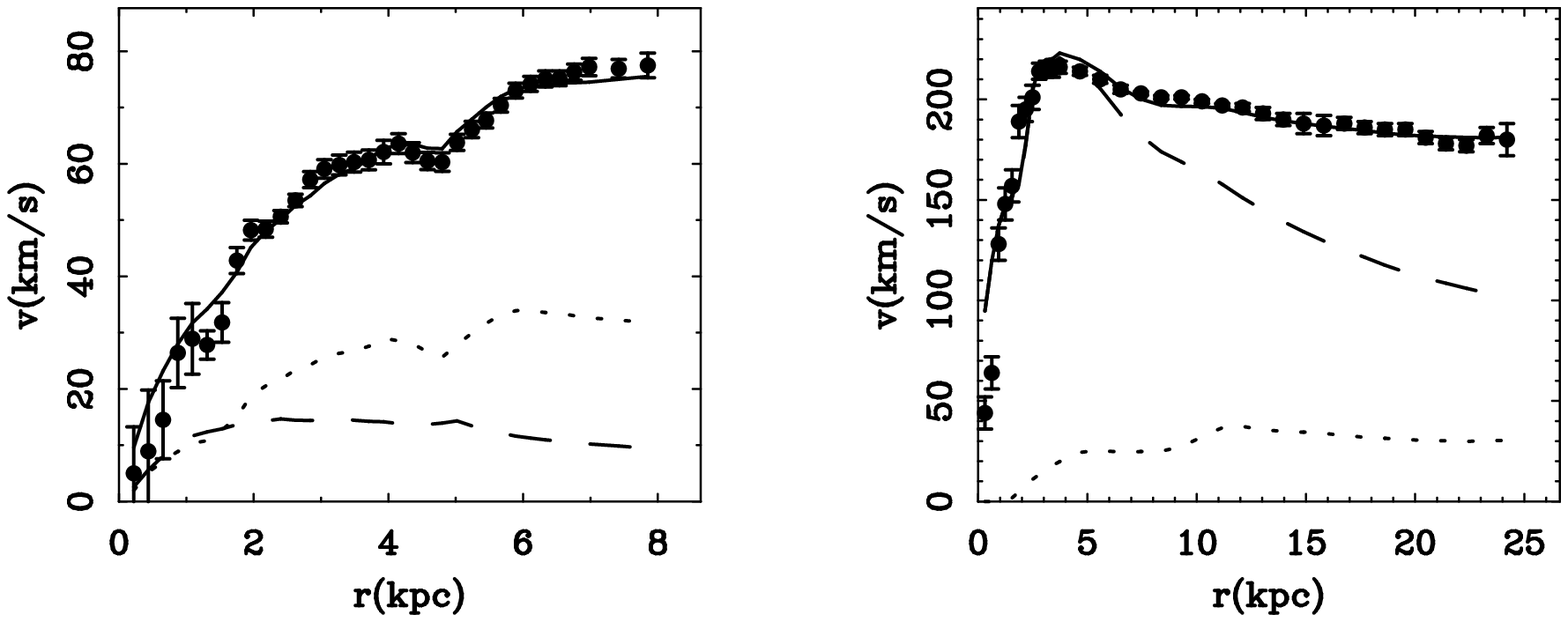}&
\includegraphics[width=0.3\columnwidth]{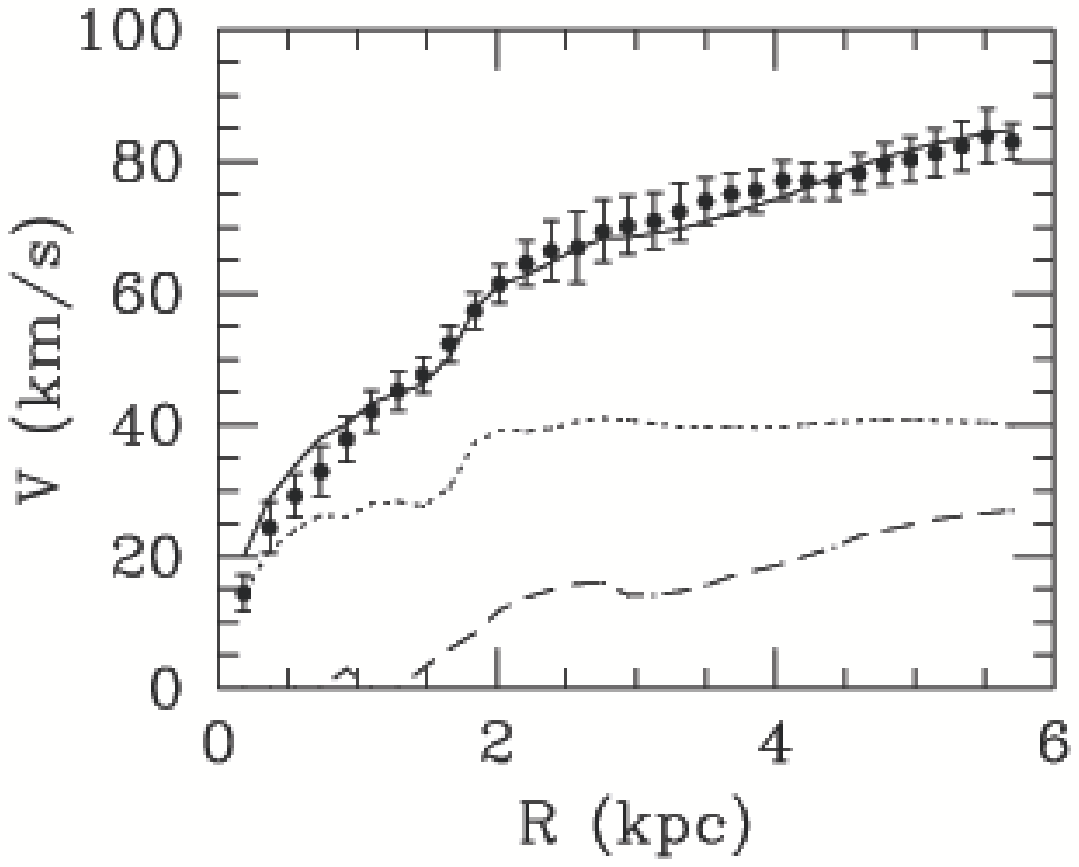}

\end{tabular}
\caption{The rotation curves of a variety of galaxies from
low-mass-low-speed galaxies that are everywhere of low-acceleration
(such as the galaxies shown at the lower right and left)
to high-mass-high-speed galaxies that have high accelerations in
the inner parts transiting to low accelerations at a finite radius
(such as the middle plot). The data is shown as points with
error bars. The line through the data is the MOND curve. Upper left
(curtesy of Bob Sanders): dashed line shows the predicted Newtonian
curve fitted to match the inner points. Lower left (from
\cite{sm02}): dotted curve is the Newtonian curve calculated for the
gas alone, and the dashed line the same for the stars (they add in
quadrature). Middle (from \cite{sm02}): dashed line the Newtonian
curve for the stars, and the dotted curve for the gas. Upper
right (from
\cite{sanders06}) and Lower right (curtesy of Bob Sanders): the
dotted line is the Newtonian curve for the stars, the dashed curve
for the gas.} \label{fig2}
\end{figure}
I show in Fig.\ref{fig2} examples of such analysis for five disc
galaxies that span the gamut of galaxy properties: from low-mass,
low-speed, gas-rich galaxies with a still rising RC, to high-mass,
high-speed, gas-poor galaxies, with a still declining RC.

\par
The MOND curves shown involve only one parameter, the mass-to-light
ratio, $M/L$, which is the conversion factor from stellar light to
mass (the gas mass is measured directly with no need for conversion).
In cases where the gas strongly dominates the mass budget, such as
for the galaxy shown on the lower left, the MOND prediction hardly
depends on the assumed $M/L$ value, and the MOND curve is
practically an absolute prediction. When the stars dominate, $M/L$
may be viewed as a parameter chosen to fit the first few points on
the rotation curve; the rest of the curve is then predicted
uniquely. Note also, that $M/L$ is not a completely free parameter:
it can be deduced for a galaxy of a given spectrum from theoretical
calculations of stellar population synthesis, within some margins.
The values required by the MOND fits were found to agree with those
theoretical values (e.g., \cite{sv98}). In DM fits one usually has
full freedom to adjust the radial scale and the normalization of the
a priori unknown contribution of the DM halo, in addition to the
stellar $M/L$ value.
\par
We see that some measured RCs exhibit features--such as a dip, or
a sudden rise--which are well reproduced by the predicted MOND
curves. These features are of prime importance in assessing the
performance of competing paradigms. They result from relatively
sharp features in the baryonic mass distribution of the disc. As
this is the only source for the MOND curve, the feature appears in
this curve in full strength (as it does in the Newtonian curve
without DM). However, adding a Dominant DM component, which does not
have thae feature (indeed, which cannot have the feature if it is
spherical) washes the feature out in the best-fit DM curve.
\par
\begin{center}
\begin{figure}
\begin{tabular}{rl}
\tabularnewline
\includegraphics[width=0.45\columnwidth]{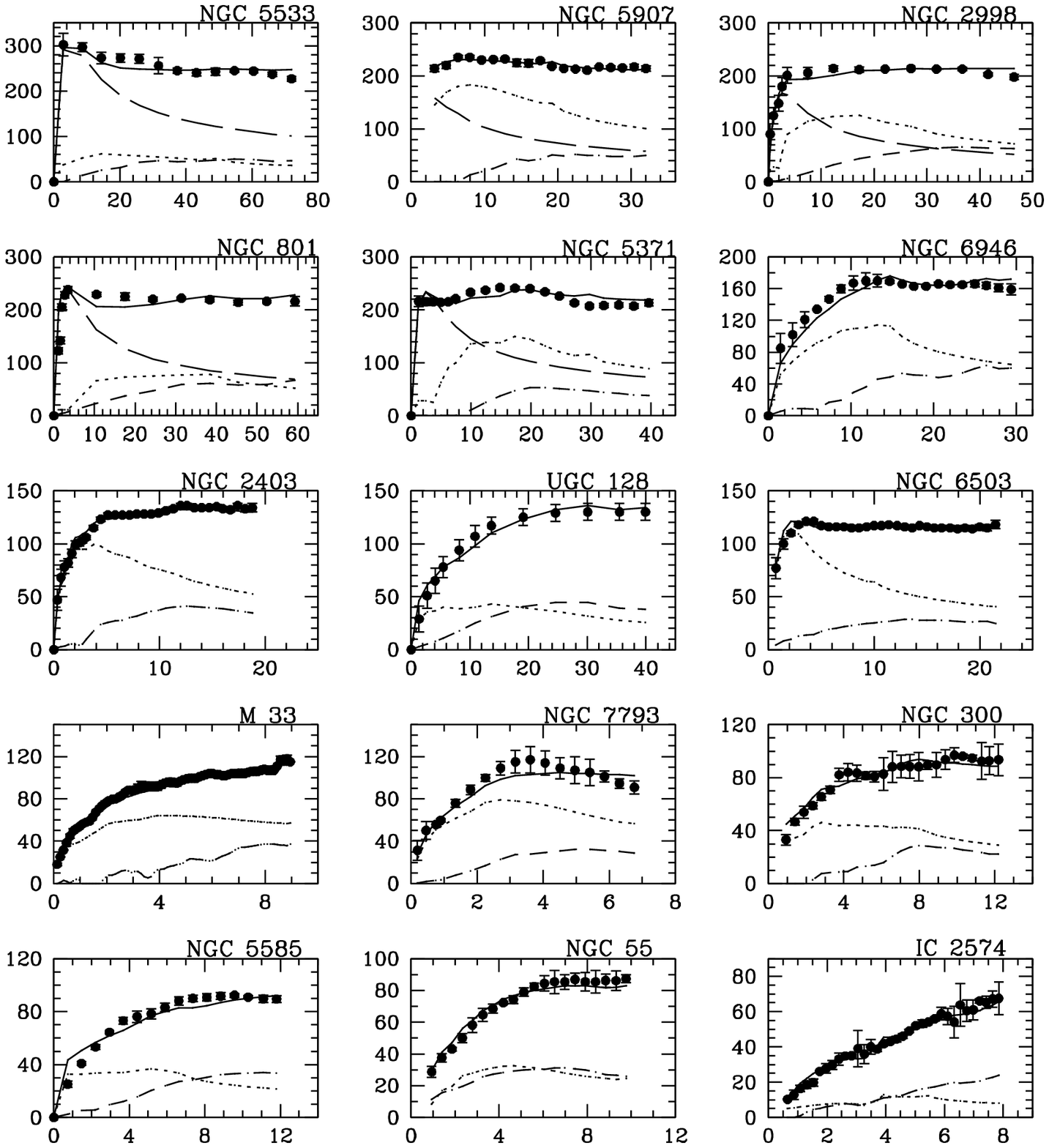} &
\includegraphics[width=0.45\columnwidth]{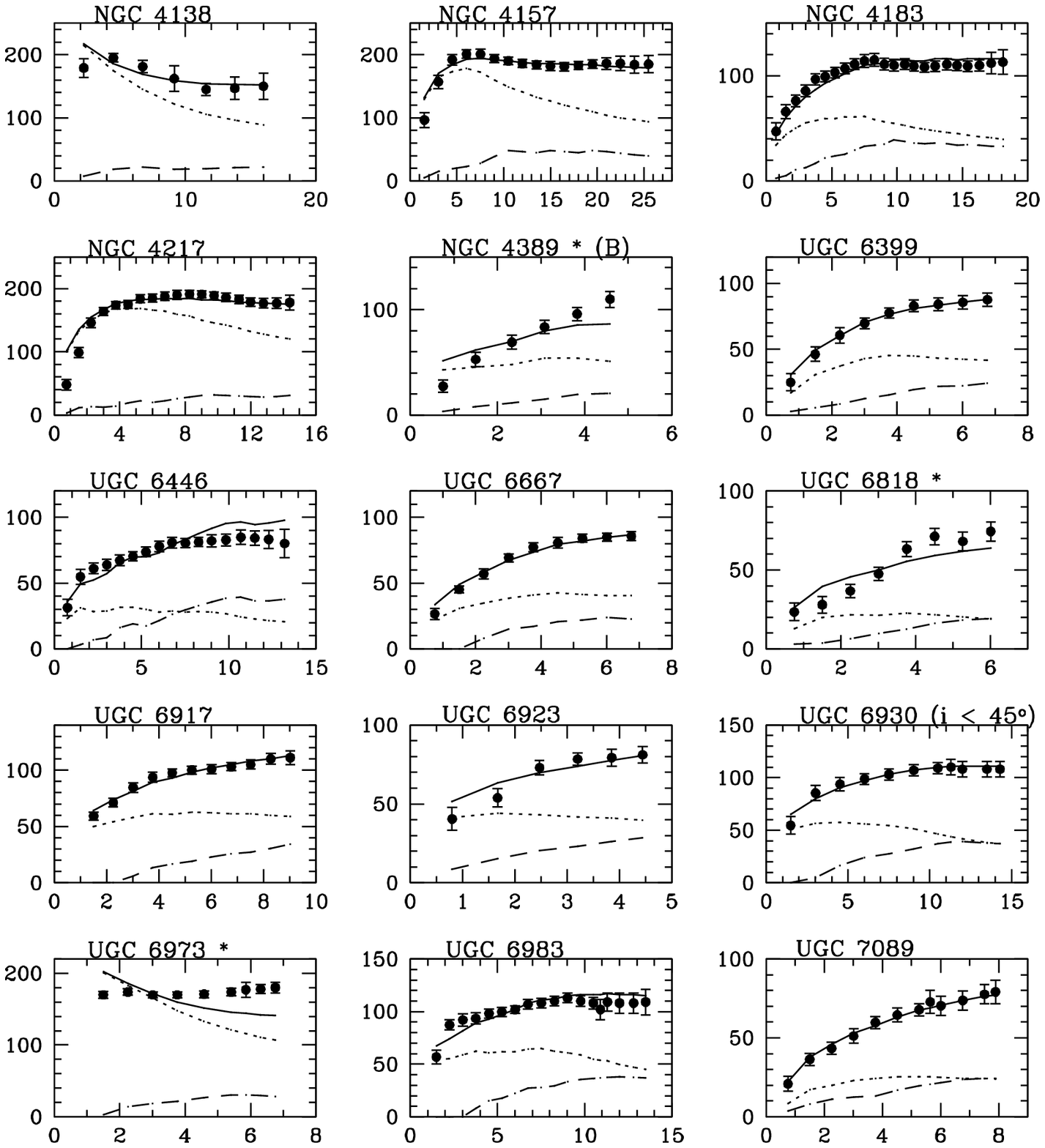}\\
\end{tabular}\par
\caption{Additional MOND rotation curves from \cite{sanders96} and
\cite{dbm98} (left) and from \cite{sv98} (right). (MOND curves in
solid; stellar disc Newtonian curves in dotted; gas in dot-dash; and
stellar bulge in long dashed.)} \label{fig3}
\end{figure}
\end{center}
\begin{figure}
\begin{tabular}{rcl}
\tabularnewline
\includegraphics[width=0.3\columnwidth]{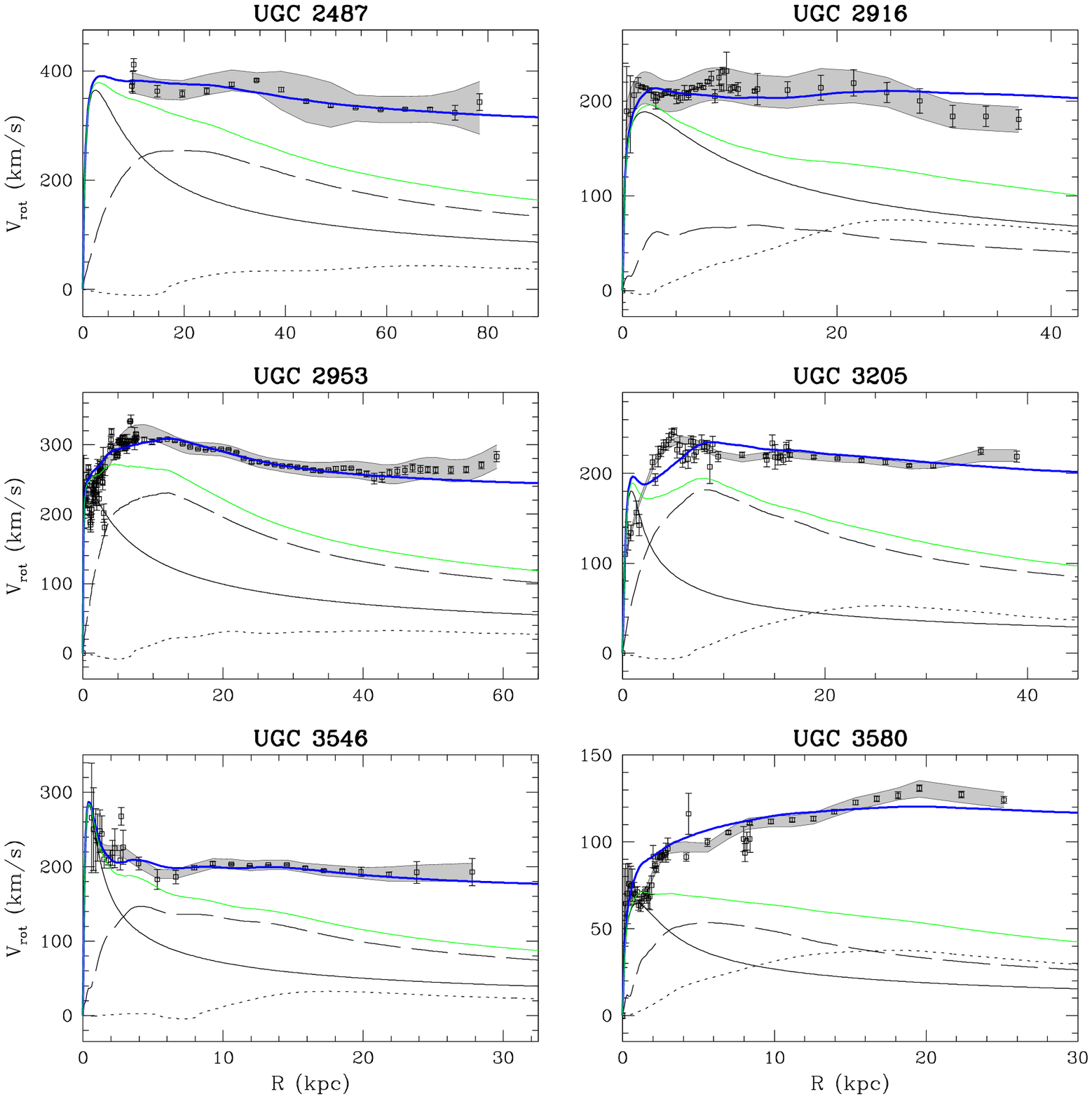} &
\includegraphics[width=0.3\columnwidth]{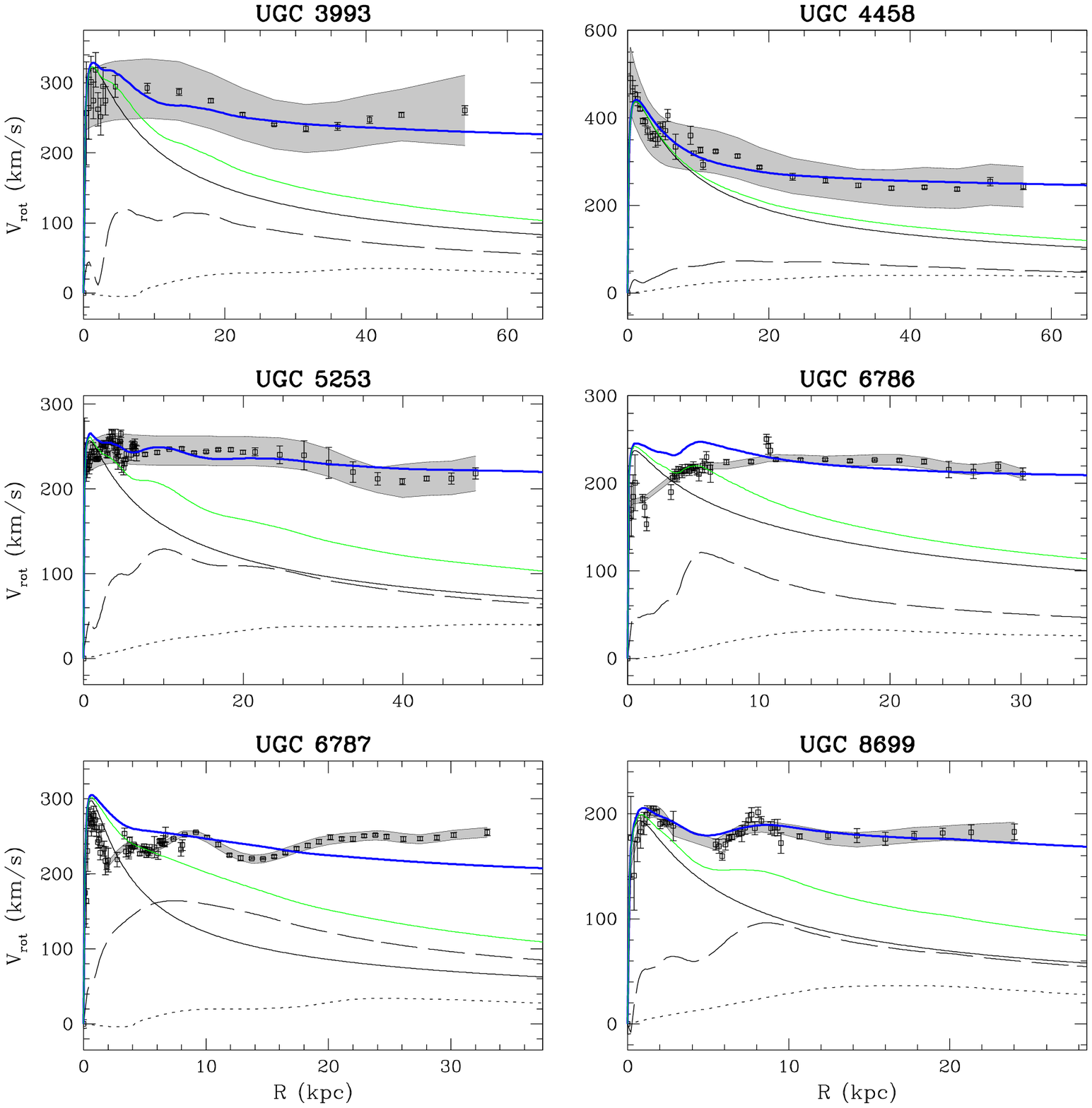} &
\includegraphics[width=0.3\columnwidth]{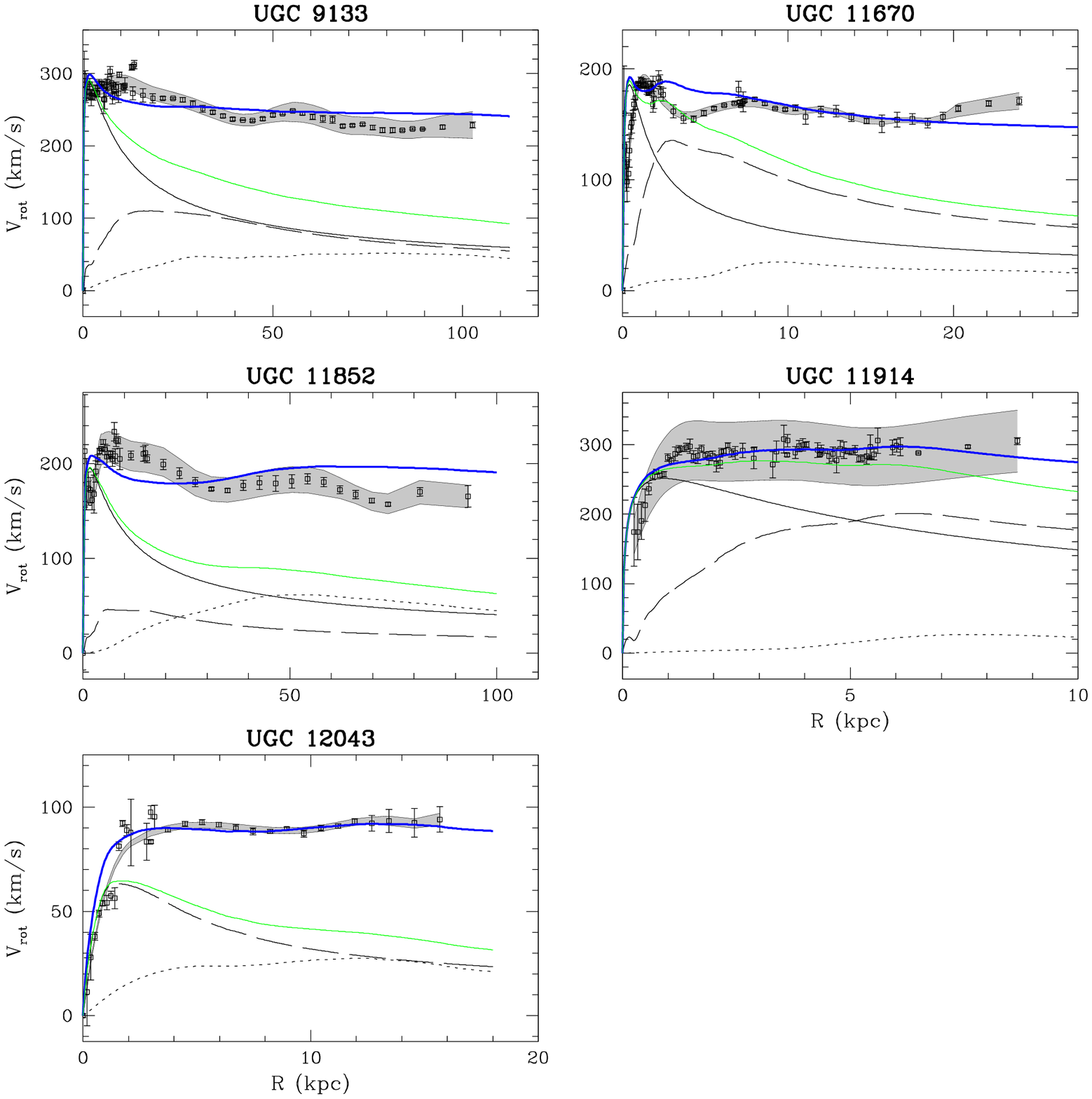}\\
\end{tabular}\par
\caption{Additional MOND rotation curves from \cite{sn07}.  The grey
shaded bands give the allowed range due to inclination
uncertainties. Thin black solid, dashed and dotted lines give the
contributions from stellar bulge, disc and gas respectively. The
thin green (grey) line gives the Newtonian sum of the individual
components and the bold blue (grey) lines gives the total MOND
rotation curve.}\label{fig4}
\end{figure}

Figures \ref{fig3}, \ref{fig4} show more results of MOND
rotation-curve analyses.

\subsection{Round systems}
Dwarf spheroidal satellites of the Milky Way are very low
acceleration systems. They indeed show large mass discrepancies, as
was predicted from MOND\cite{mil83} before these discrepancies were
measured. The state of the art analysis of these systems and
references to earlier work can be found in \cite{angus08,serra09},
where it is found that, with one or two exceptions, perhaps, MOND
indeed accounts for the large mass discrepancies deduced with
Newtonian dynamics.
\par
As has been realized through many analyses starting in the early
1990s, the only system type where MOND has failed systematically to
account for the full mass discrepancy comprise galaxy clusters. The
situation has been reviewed recently in \cite{mil08a} (and many
references therein). Briefly: the typical global mass discrepancy in
clusters (within say 2 Mpc) is about a factor 7-10). MOND corrects
this by reducing the required mass by a factor 3-5, but leaves still
a factor of two discrepancy. So roughly as much `dark mass' is still
required as there is presently visible baryonic mass.  MOND
adherents have attributed this dark mass to possibly neutrinos or to
some form of yet undetected cluster baryonic matter (CBDM). The
distribution of the CBDM in the cluster is rather like that of the
galaxies, being rather more centrally concentrated than the hot gas,
which is makes the lion's share of the visible component. This makes
the mass discrepancy that remains after the MOND correction rather
stronger in the cluster core (within a few hundred kpc). The
appearance of the post-collision `bullet cluster', which has been
bruited as `direct proof of the existence of DM', conforms exactly
to what is expected from this picture, long known from studies of
individual clusters. It only reiterates the fact that a still
undetected mass component exists in cluster cores, but does not
pinpoint it as the putative particle DM.

\subsection{The solar system}
We know of two concrete solar system phenomena that MOND could
impact. The first is the so called Pioneer anomaly: a yet
unexplained constant acceleration of the two Pioneer spacecraft
towards the sun, of a magnitude $\approx \baz\equiv 2\pi \az$. While
this anomaly may yet turn out to result from a mundane,
underestimated systematics, and not from new physics, MOND does have
the potential to explain such an anomaly. Because a similar anomaly
in the motions of the planets can be ruled out by a large margin, a
new-physics explanation must hinge on the differences between the
planetary orbits (bound, nearly circular) and those of the
spacecraft (unbound, highly hyperbolic) (see \cite{mil08} and
references therein for details).
\par
I have recently pointed out \cite{mil09c} that certain MOND theories
predict an unavoidable MOND effect in the inner solar system
(planetary zone) having to do with the galactic acceleration field
in which the solar system is falling (the external-field effect of
MOND). Near the sun--i.e. at distances much smaller than the MOND
radius of the sun, $R\_M\equiv(\msun G/\az)^{1/2}\approx 8\times
10^3 {\rm a.u.}$--this appears as an anomalous quadrupole field with
anomalous potential $\f\_{an}=(q/4)\az R\_M^{-1}(2z^2-x^2-y^2)$.
Here, the $z$-axis is in the direction of the galactic center, and
$q\sim -0.1$ depends on the strength of the galactic acceleration
field at the position of the sun, and on the form of the MOND
extrapolating function of the theory. This anomaly causes very
distinctive effects on planetary motions, such as an anomalous
perihelion precession with characteristic dependence on the length
of the semi-major axis and on its orientation relative to the
galactic center's direction. This effect seems to be below present
day detection capabilities, but not inaccessibly below.

\section{MOND vs. cold dark matter}
MOND and DM were initially advanced to explain similar phenomena:
the mass discrepancies in galactic systems. This may create the
impression that they are twin paradigms, that the deductions we make
from both are of similar force, and that the only way to decide
between them is to compare the performance of their predictions
against the data in cases where the predictions differ. This is far
from being the case.
\par
I concentrate henceforth on cold dark matter (CDM), the eminently
favored version of the DM paradigm today. Deductions made in CDM and
in MOND, pertaining even to the same phenomena and measurements,
such as galaxy rotation curves, the existence and nature of a TF
relation, the distribution of the mass discrepancy in a given
system, etc., are of a totally different nature. In MOND, these are
strict and inevitable predictions of all aspects of the mass
discrepancy in any given object, based only on the distribution of
normal (baryonic) matter as it is now observed. Such predictions can
be made for any given, individual system. In the CDM paradigm, all
such deductions for a given object would depend strongly on its
detailed history: initial collapse, merger history, cannibalism of
satellites, gas accretion, star formation, baryon ejection by
supernova explosions and ram pressure effects, dissipation,  angular
momentum exchange, cooling and heating, etc., etc.--details that we
do not know, and cannot know for any individual object, with very
few, important exceptions (see below). To make matters worse, the
two major components, baryons and the putative, weakly interacting
CDM, partake very differently in all the above processes. Despite
many works in the literature that attempt to account for  these via
so called `semi analytic' calculations, the bare fact is that CDM is
incapable of truly predicting the mass discrepancies in galactic
systems (again with that important exception to be discussed below).
CDM is utterly incapable of predicting even the most basic number:
the baryon fraction in a given galaxy.\footnote{We know that this
fraction in a galaxy today is typically much smaller than the cosmic
fraction with which proto-galaxies must have started their life. So
most of the baryons must have been lost on the way, somehow (in the
CDM paradigm), exactly how, and what small fraction of it is left,
are the grist for `semi analytic' models.} But such numbers are the
preliminaries for arriving at a deduced TF relation for example, to
say nothing of predicting full rotation curves. What we then see
often as CDM analysis of rotation curves are merely best fits, which
assume, in addition to the baryons, a certain DM halo mass
distribution with the total mass and radial scale of the
distribution left as free parameters. The MOND rotation curves have
non of this. At worst they involve one free parameter for converting
stellar light into baryonic mass (which CDM fits also have in
addition to the other parameters). But even this is not needed in
many instances, as explained above.
\par
The other important point to note is that even if CDM is one day
found capable of deducing from first principles some statistical,
population properties pertaining to the relations between baryons
and DM (such as the TF relation), it it highly unlikely to ever make
predictions for individual systems, as MOND does.
\par
To recapitulate, the fact that galaxies follow closely the MOND
predictions argue against CDM in two ways. In the first place
because they show that there is a paradigm that can predict all
those observed regularities that CDM cannot predict. But, in
addition, the fact that such strict regularities exist at all--even
irrespective of the existence of MOND--argues against CDM, because
the relations between the baryonic and DM components is expected to
be highly haphazard in the CDM paradigm.
\par
I mentioned above that there is one important exception to CDM's
incapability of predicting baryon-to-DM relations. This concern so
called tidal dwarfs: the small `phoenix' galaxies that are born from
the gaseous tidal tails spewed in the aftermath of high speed
collisions between ready made galaxies. Such tidal dwarfs are an
exception to the above rule because their history is brief and
involves only a few, rather well understood processes. Simulations
show that hardly any of the CDM purportedly surrounding the two
colliding, parent galaxies makes its way into the the tidal tails
and the tidal dwarfs that form in them. CDM then predicts no mass
discrepancies to speak of in tidal dwarfs. In MOND, the criterion
for the appearance of a mass discrepancy is only the accelerations
involved, not, in any way, the process of its formation.
\par
So, what is known about mass discrepancies in tidal dwarfs? There
are now two possible relevant instances. In the first, Bournaud et
al. \cite{bournaud07} have identified and analyzed three tidal
dwarfs in the tidal debris around NGC 5291. They find, in contrast
with the expectations from the CDM paradigm, that the three dwarfs
do exhibit mass discrepancies of about a factor three. Subsequent
analysis \cite{mil07,gentile07} showed that these agree well with
the predictions of MOND for these systems. In another possible
instance, it has been argued recently \cite{metz09} that most of the
dwarf spheroidal satellites of the Milky Way are remnant tidal
dwarfs in a single historical tidal tail. This deduction is based on
the fact that both the position of these dwarfs and their motions
indicate that they live and move in a disc around the Milky Way.
This, and other aspects of the population, are difficult to
understand unless, it is argued, they formed as tidal dwarfs. If so,
the CDM paradigm implies they should not exhibit the large mass
discrepancies they are known to show. In MOND, as said above, we do
expect them to show large discrepancies because they are very law
acceleration systems.
\par
We saw that even the most common tests, such as RC analysis, hold a
strong potential for deciding between MOND and CDM, for those who
are willing to grasp better the differences in the nature of these
deductions in MOND and CDM. However, in addition there are quite a
few phenomena  on which MOND and CDM make disparate predictions. I
already discussed above the case in point concerning tidal dwarfs. A
few other examples are: 1. MOND, but not CDM, predicts that in disc
galaxies there will be a strong disc component of `phantom' DM in
addition to the extended spheroidal one. The predicted disc
component appears only where the baryonic disc is, and only where
the orbital accelerations in the disc are small in the MOND sense
\cite{mil01}. 2. Insistence on interpreting MOND results with DM
will lead to negative DM densities in some locations \cite{mil86b}.
3. We saw already that some MOND theories predict an anomalous
quadrupole field in the inner solar system, which would be quite
unexpected from CDM.
\par
To summarize this section: The success of MOND predictions implies
that baryons alone strictly determine the acceleration fields of
galactic objects. This conflicts with the expectations in the CDM
paradigm (where these fields are, by and large, dominated by the
contribution of the DM) because of the haphazard formation and
evolution of galactic objects, and because baryons and DM are given
to very different behaviors during the evolution. This is evinced by
the very different characteristics of the baryonic and the putative
DM components in galaxies today (e.g., the very small baryon-to-DM
fraction in galaxies compared with the cosmic value, the highly
concentrated, and oftentimes disc-like, baryonic component compared
with the much more extended, spheroidal DM component). It is thus
highly unlikely that DM will someday reproduce MOND.

\section{The cosmological connection}
So far I discussed  MOND only from the phenomenological point of
view: It has been advanced to obviate the need for DM, and it has,
indeed, managed to predict, explain, and organize large amounts of
observations in individual systems, and of population laws and
regularities, without invoking DM. However, there is an additional
aspect revealed by MOND that may hint at more profound implications:
The value of the MOND constant $\az$, as determined from the various
laws listed above, is $\az\approx 1.2\times 10^{-8}\cmss$. This
value is tantalizingly close to acceleration parameters that
characterize cosmology \cite{mil83,mil99}. Define, more
conveniently, $\baz\equiv 2\pi\az$, then one finds
 \beq \baz\approx \aH=c H_0,~~~~~~~~~~
\baz\approx \aL=c(\Lambda/3)^{1/2}.  \eeqno{cosmo}
 The first acceleration parameter, $\aH$, is associated with the
 present day expansion
rate $H_0$, while the second is defined by the present density of
`dark energy' as determined by $\Lambda$. The mysterious fact that
today the two cosmic acceleration parameters are nearly the same
defines the `cosmic coincidence' puzzle (another coincidence puzzle
being the proximity of the baryon and DM densities, when the two
components are thought to have formed by totally different and
unrelated processes). The fact that $\baz$--an acceleration that
emerges very clearly and forcibly from the phenomenology of galactic
systems--also nearly equals these two cosmic accelerations should be
viewed, in my opinion, as an added cosmic coincidence.
\par
Much has been said about the significance of this last `coincidence'
and its possible origins and implications (e.g., in
\cite{mil83,sanders97,mil99,blt08,blt09,mil09a,mil09e}). Here I only
touch briefly on a recent observation that might connect MOND with
cosmology even more firmly \cite{mil09a}. If the `dark energy' is a
cosmological constant (CC), then our universe is nearly of a de
Sitter (dS) geometry in the vicinity of the present time. In the
future, as the CC becomes increasingly dominant over matter, the
geometry will tend towards exact dS geometry. Now, the second
coincidence in eq.(\ref{cosmo}) is tantamount to the MOND
acceleration being related to the radius, $\ell$, of this asymptotic
dS cosmos by $\baz\approx c^2/\ell$. The possible tighter ties of
MOND with cosmology may then be reflected in the symmetries that
characterize these two structures. The symmetry (isometry) group of
a 4-D dS space-time is the group of rotations in Minkowski 5-D space
time; i.e. $SO(4,1)$. If MOND, and in particular the deep MOND
limit, is in some way a reflection of the near dS nature of our
cosmos, the symmetries characterizing MOND might reflect the
symmetries of the dS space-time, or perhaps both symmetries are a
result of the symmetries of the theory that underlies both. Our
actual universe is not quite of dS geometry; so we do not expect
MOND at large to share symmetries with that geometry. Indeed MOND
with its full coverage is not known to have symmetries beyond the
standard rotation, and space- and time-translation invariance. But
the deep MOND limit itself does have added symmetries in certain
formulations. For example, the deep MOND limit of the nonlinear
Poisson formulation eq.(\ref{poisson}) has been found to have the
full conformal symmetry in three space dimensions. Together with the
space rotations and translations these conformal symmetries form a
larger group which is equivalent to the SO(4,1) geometrical symmetry
group of exact dS space-time. This has been proposed \cite{mil09a}
as a possible indication of a connection similar to the celebrated
duality of conformal field theories on the boundary of an Anti-dS
(or dS) space-time and gravity in that space-time. Here as well, the
space on which the extended deep-MOND theories act is the Euclidean
three-dimensional space (or rather its compactification to a
Euclidean 3-D sphere), and this is indeed the past (or future)
boundary of a dS space-time.

\section{Open questions}
Beside the yet unexplained remaining factor-of-two discrepancy in
galaxy clusters, the major challenge to MOND is to account fully for
the need for cosmological DM. In other words we need a MOND inspired
mechanism that will account for the effects in galaxy formation,
CMB, etc., of what in standard dynamics is attributed to
cosmological DM with a contribution of about 0.2 to $\Omega$. One
effect of cosmological DM is to enhance structure formation because,
being neutral, it would have started to collapse before
recombination. MOND clearly has the ability to enhance gravitational
collapse over standard dynamics without DM. Several studies of
structure formation in MOND inspired schemes, some using somewhat
crude NR models, some using relativistic theories such as TeVeS,
have been conducted, and have shown such enhancing effects: there
are indeed aspects of MOND that could replace cosmological DM
effects \cite{sanders01,nusser02,kutschera02,knebe04,
skordisetal06,dodelson06,zfs08,ll08}. It has to be realized,
however, that the theory of structure formation, interaction of
matter with the CMB, etc. in MOND is still much less developed and
focussed then in standard dynamics. In my opinion a satisfactory
understanding of all these will come only after we better understand
the nature of the MOND-cosmology connection discussed above, and
when we then have a full fledged relativistic version of MOND,
which, arguably, we are still lacking.
\par
In any event, one should certainly not be dismayed by the tasks
still left to be accomplished. The very simple MOND idea proposed a
quarter of a century ago, has already achieved much more then could
be expected of it at the time of its inception. It can in fact be
argued that the richness and variety, and sheer quantity of
successful predictions in the context of galaxy dynamics is many
fold what is still required in cosmology. Understanding a single
galaxy is not as important as understanding cosmology. But,
epistemologically, the detailed prediction of a field of a single
galaxy--accounting exactly for what otherwise would result from a
detailed amount and distribution of DM--is, in my opinion, a feat on
a par with predicting the few statistical properties attributed
conventionally to cosmological DM. After all, Newtonian dynamics was
based on the observations of only one solar system.
\par
All in all, with the encouragement from past successes and the
attraction of remaining challenges, times are interesting for those
who choose to work on MOND.


\begin{theacknowledgments}
 This research was supported by a
center of excellence grant from the Israel Science Foundation.

\end{theacknowledgments}


\begin{thebibliography}{9}
\bibitem[ ()]{angus08}G. W. Angus, \emph{MNRAS} \textbf{387}, 1481
(2008)
\bibitem[ ()]{banados09}M. Ba\~{n}ados, P.G. Ferreira, and
C. Skordis, \emph{Phys. Rev. D} \textbf{79}, 063511 (2009).
\bibitem[ ()]{barnes07}E. I. Barnes, A. Kosowsky, and J. A. Sellwood,
\emph{AJ} \textbf{133}, 1698  (2007).
\bibitem[ ()]{bbs91}K. G. Begeman, A. H. Broeils, and R. H. Sanders,
 \emph{MNRAS} \textbf{249}, 523 (1991).
\bibitem[ ()]{begum04}A. Begum, and J. Chengalur,
 \emph{AA} \textbf{413}, 525 (2004).
\bibitem[ ()]{bek04} J.D. Bekenstein, \emph{Phys. Rev. D} \textbf{70},
083509 (2004).

\bibitem[ ()]{bek06} J.D. Bekenstein, \emph{Contemp. Phys.} \textbf{47},
387 (2006).
\bibitem[ ()]{bm06}J. Bekenstein, and J. Magueijo, \emph{Phys. Rev. D}
\textbf{73},
103513 (2006).
\bibitem[ ()]{bm84}J. Bekenstein, and M. Milgrom, \emph{ApJ} \textbf{286},
7 (1984).
\bibitem[ ()]{bl07}L. Blanchet,
\emph{Class. Quant.Grav.} \textbf{24}, 3541 (2007).
\bibitem[ ()]{blt08}L. Blanchet, and A.  Le Tiec,
\emph{Phys. Rev. D} \textbf{78}, 024031 (2008).
\bibitem[ ()]{blt09}L. Blanchet, and A.  Le Tiec,
\emph{Phys. Rev. D} \textbf{80}, 023524 (2009).
\bibitem[ ()]{bdg06}D. Blas, C. Deffayet,
and J. Garriga, \emph{Class. Quant. Grav.} \textbf{23}, 1697 (2006).
\bibitem[ ()]{bdg07}D. Blas, C. Deffayet,
and J. Garriga, \emph{Phys. Rev. D} \textbf{76}, 104036 (2007).
\bibitem[ ()]{dbm98}W. J. G. de Blok, and S. S.
McGaugh, \emph{ApJ} \textbf{508}, 132 (1998).
\bibitem[ ()]{bottema02}R. Bottema, J. L. G. Pesta\~{n}a,
B. Rothberg, and R. H. Sanders, \emph{AA} \textbf{393}, 453 (2002).
\bibitem[ ()]{boulanger01}N. Boulanger, T. Damour, L. Gualtieri, and M.
Henneaux \emph{Nucl.Phys. B} \textbf{597}, 127 (2001)
\bibitem[ ()]{bournaud07}F. Bournaud, P.-A. Duc, E.
Brinks, M.  Boquien, M. Amram, U. Lisenfeld, B. S. Koribalski, F.
Walter, and V. Charmandaris, \emph{Science} \textbf{316}, 1093
(2007).
\bibitem[ ()]{brada99}R. Brada, and M. Milgrom, \emph{ApJ} \textbf{519},
 590 (1999).
\bibitem[ ()]{brada00a}R. Brada, and M. Milgrom, \emph{ApJ} \textbf{541},
 556
(2000).
\bibitem[ ()]{brada00b}R. Brada, and M. Milgrom, \emph{ApJ}
\textbf{531}, L21 (2000).
\bibitem[ ()]{bruneton07}J.-P. Bruneton, and G. Esposito-Farese,
\emph{Phys. Rev. D} \textbf{76}, 124012 (2007).
\bibitem[ ()]{cb04}L. Ciotti, and J. Binney, \emph{MNRAS} \textbf{351},
285
(2004).
\bibitem[ ()]{contaldi08}C. R. Contaldi, T. Wiseman, and B. Withers,
\emph{Phys. Rev. D} \textbf{78}, 044034 (2008).
\bibitem[ ()]{corbelli07}E. Corbelli, P.
Salucci, \emph{MNRAS}, \textbf{374}, 1051 (2007).
\bibitem[ ()]{dms08}D. C. Dai, R. Matsuo, and G. Starkman,
\emph{arXiv:0811.1565}
\bibitem[ ()]{dodelson06}S. Dodelson, and M. Liguori,
\emph{Phys. Rev. Lett.},
 \textbf{97}, 231301 (2006).
\bibitem[ ()]{donato09}F. Donato, G. Gentile, P. Salucci,
C. Frigerio Martins, M. I. Wilkinson, G. Gilmore, E.K. Grebel, A.
Koch, and R. Wyse, \emph{MNRAS}  \textbf{397}, 1169 (2009).
\bibitem[ ()]{fbzh07}B. Famaey, J.-P. Bruneton, and
H. S. Zhao, \emph{MNRAS} \textbf{377}, L79 (2007).

\bibitem[ ()]{gentile04}G. Gentile, P. Salucci, U. Klein, D. Vergani,
and P. Kalberla, \emph{MNRAS} \textbf{351}, 903 (2004).
\bibitem[ ()]{gentile07}G. Gentile, B. Famaey, F.
Combes, P. Kroupa, H. S. Zhao, and O. Tiret, \emph{AA} \textbf{472},
L25 (2007).
\bibitem[ ()]{gentile07a}G. Gentile, P. Salucci, U. Klein, and
G. L. Granato,
 \emph{MNRAS} \textbf{375}, 199 (2007).
\bibitem[ ()]{gentile09}G. Gentile, B. Famaey, H. S. Zhao, and P. Salucci,
 \emph{Nature} \textbf{461}, 627, (2009).
\bibitem[ ()]{graham06}A. W. Graham, D. Merritt,
B. Moore, J. Diemand, and B. Terzi\'c, \emph{AJ} \textbf{132}, 2711
(2006).
\bibitem[ ()]{kalberla07}P.M.W. Kalberla, L. Dedes, J. Kerp, and U. Haud
\emph{AA} \textbf{469}, 511 (2007).
\bibitem[ ()]{kent87}S. M. Kent, \emph{AJ} \textbf{93}, 816 (1987).
\bibitem[ ()]{knebe04}A. Knebe, and B. K. Gibson, \emph{MNRAS}
\textbf{347}, 1055 (2004).
\bibitem[ ()]{ll08}C. Llinares, A. Knebe, and
H. S. Zhao, \emph{MNRAS} \textbf{391}, 1778 (2008).
\bibitem[ ()]{mcgaugh98}S. S. McGaugh, and W. J. G. de Blok,
\emph{ApJ} \textbf{499}, 66 (1998).
\bibitem[ ()]{mcg05}McGaugh, S.S., 2005b, Proceedings of the XXIst IAP Colloquium
"Mass Profiles and Shapes of Cosmological Structures", Eds. G.
Mamon, F. Combes, C. Deffayet, B. Fort, EDP Sciences;
arXiv:astro-ph/0510620.

\bibitem[ ()]{mcgaugh00}S. S. McGaugh, J. M. Schombert, G. D. Bothun,
and W. J. G. de
Blok, \emph{ApJ} \textbf{523}, L99 (2000).
\bibitem[ ()]{metz09}M. Metz, P.  Kroupa, and H. Jerjen,
\emph{MNRAS} \textbf{394}, 2223
(2009).
\bibitem[ ()]{mil83}M. Milgrom, \emph{ApJ} \textbf{270}, 365 (1983).
\bibitem[ ()]{mil86a}M. Milgrom, \emph{ApJ} \textbf{302}, 617
(1986).
\bibitem[ ()]{mil86b}M. Milgrom, \emph{ApJ} \textbf{306}, 9
(1986).
\bibitem[ ()]{mil88}M. Milgrom, \emph{ApJ} \textbf{333}, 689 (1988).

\bibitem[ ()]{mil94}M. Milgrom, \emph{Ann. Phys.} \textbf{229}, 384
(1994).
\bibitem[ ()]{mil97}M. Milgrom,  \emph{Phys. Rev. E} \textbf{56}, 1148
(1997).
\bibitem[ ()]{mil99}M. Milgrom,  \emph{Phys. Lett. A} \textbf{253}, 273
(1999).
 \bibitem[ ()]{mil01} M. Milgrom, \emph{MNRAS} \textbf{326}, 1261 (2001).
\bibitem[ ()]{mil02}M. Milgrom,   \emph{J. Phys. A} \textbf{35}, 1437
(2002).
\bibitem[ ()]{mil07} M. Milgrom, \emph{ApJL} \textbf{667}, L45 (2007).
\bibitem[ ()]{mil08a}M. Milgrom, \emph{New Astron.Rev.} \textbf{51}, 906
(2008).
\bibitem[ ()]{mil08} M. Milgrom,   In Proceedings XIX Rencontres de
Blois; arXiv:0801.3133 (2008).
\bibitem[ ()]{mil09a} M. Milgrom, \emph{ApJ} \textbf{698}, 1630
(2009).
\bibitem[ ()]{mil09b} M. Milgrom, \emph{MNRAS} \textbf{398}, 1023 (2009).
\bibitem[ ()]{mil09c} M. Milgrom, \emph{MNRAS} \textbf{399}, 474 (2009).
\bibitem[ ()]{mil09d} M. Milgrom, \emph{MNRAS} \textbf{403}, 886 (2010),
arXiv:0911.5464.
\bibitem[ ()]{mil09e} M. Milgrom, \emph{Phys. Rev. D} \textbf{80}, 123536,
arXiv:0912.0790 (2009).
\bibitem[ ()]{mil10a} M. Milgrom, \emph{MNRAS} \textbf{405}, 1129 (2010),
arXiv:1001.4444.
\bibitem[ ()]{ms07}M. Milgrom, and R.H. Sanders, \emph{ApJ}
\textbf{658}, L17 (2007).
\bibitem[ ()]{nlc07a}C. Nipoti, P. Londrillo, and L. Ciotti, \emph{MNRAS}
 \textbf{381}, L104 (2007).
\bibitem[ ()]{nlc07b}C. Nipoti, P. Londrillo, and L. Ciotti, \emph{ApJ}
 \textbf{660}, 256 (2007).
\bibitem[ ()]{nip08}C. Nipoti, L. Ciotti, J. Binney, and P. Londrillo,
\emph{MNRAS} \textbf{386}, 2194 (2008).
\bibitem[ ()]{nusser02}A. Nusser, \emph{MNRAS} \textbf{331}, 909, (2002).
\bibitem[ ()]{sagi09}E. Sagi, \emph{Phys. Rev. D}
\textbf{80}, 044032 (2009).
\bibitem[ ()]{sagi08}E. Sagi, and J. D. Bekenstein, \emph{Phys. Rev. D}
\textbf{77}, 024010 (2008).
\bibitem[ ()]{sanders96}R. H. Sanders, \emph{ApJ} \textbf{473}, 117 (1996).
\bibitem[ ()]{sanders97}R. H. Sanders, \emph{ApJ} \textbf{480}, 492 (1997).
\bibitem[ ()]{sanders01}R. H. Sanders, \emph{ApJ} \textbf{560}, 1 (2001).
\bibitem[ ()]{sanders05}R. H. Sanders, \emph{MNRAS} \textbf{363}, 459
(2005).
\bibitem[ ()]{sanders06}R. H. Sanders , Third Aegean Summer School,
 \emph{The Invisible Universe: Dark Matter and Dark Energy},
arXiv:astro-ph/0601431 (2006).
\bibitem[ ()]{sm02}R. H. Sanders, and S. S. McGaugh, \emph{ARAA}
 \textbf{40},
263 (2002).
\bibitem[ ()]{sn07}R. H. Sanders, and E. Noordermeer, \emph{MNRAS}
\textbf{379},
702 (2007).
\bibitem[ ()]{sv98}R. H. Sanders, and M. A. W. Verheijen,
\emph{ApJ} \textbf{503}, 97 (1998).
\bibitem[ ()]{scarpa06}R. Scarpa, ``Modified Newtonian Dynamics,
an Introductory Review'' in \emph{First
crisis in cosmology}, edited by E.J. Lerner, J.B. Almeida, AIP
Conference
 822, American Institute of Physics, New York, 2006, pp. 253
\bibitem[ ()]{serra09} A. L. Serra, G. W. Angus, and A. Diaferio,
arXiv:0907.3691
\bibitem[ ()]{skordis06}C. Skordis, \emph{Phys. Rev. D} \textbf{74},
 103513 (2006).
\bibitem[ ()]{skordis08}C. Skordis, \emph{Phys. Rev. D} \textbf{77},
 123502 (2008).
\bibitem[ ()]{skordis09}C. Skordis, \emph{Class. Quant. Grav.}
\textbf{26 (14)}, 143001 (2009).
\bibitem[ ()]{skordisetal06}C. Skordis, D. F. Mota, P. G. Ferreira, and
C.  Boehm, \emph{Phys. Rev. Lett.} \textbf{96}, 011301 (2006).
\bibitem[ ()]{kutschera02}S. Stachniewicz, and M. Kutschera,
\emph{Acta physica Polonica} \textbf{B32}, 362, (2002).
\bibitem[ ()]{stark09}D. V. Stark, S. S. McGaugh, and R.A.
 Swaters, \emph{AJ} \textbf{138}, 392 (2009).
\bibitem[ ()]{tiret07}O. Tiret, and F. Combes,
in \emph{Formation and Evolution of Galaxy Disks}, edited by J. G.
Funes, and E. M. Corsini, arXiv:0712.1459 (2007).
\bibitem[ ()]{tiret08}O. Tiret, and F. Combes, \emph{AA} \textbf{483},
719 (2008).
\bibitem[ ()]{trachternach09}C. Trachternach, W. J. G. de Blok, S. S.
 McGaugh,  J. M. van der Hulst, and R. -J. Dettmar, arXiv:0907.5533
  (2009).
\bibitem[ ()]{wang08}Y. Wang, X. Wu, and H. S. Zhao,
\emph{ApJ} \textbf{677}, 1033 (2008).
\bibitem[ ()]{wu07} X. Wu, H. S. Zhao, B. Famaey, G. Gentile,
O. Tiret, F. Combes, G. W. Angus, and A. C. Robin,  \emph{ApJ}
\textbf{665}, L101 (2007).
\bibitem[ ()]{wu09} X. Wu, H. S. Zhao, Y. Wang, C. Llinares, and A.
Knebe, \emph{MNRAS} \textbf{396}, 109 (2009).
\bibitem[ ()]{zfs06}T. G. Zlosnik, P. G. Ferreira, and G. D. Starkman,
\emph{Phys. Rev. D} \textbf{74}, 044037 (2006).
\bibitem[ ()]{zfs07}T. G. Zlosnik, P. G. Ferreira, and G. D. Starkman,
\emph{Phys. Rev. D} \textbf{75}, 044017 (2007).
\bibitem[ ()]{zfs08}T. G. Zlosnik, P. G. Ferreira, and G. D. Starkman,
\emph{Phys. Rev. D} \textbf{77}, 084010 (2008).

\end{thebibliography}
\end{document}